\documentclass[conference,letterpaper]{IEEEtran}

\usepackage[utf8]{inputenc} 
\usepackage[T1]{fontenc}
\usepackage{url}
\usepackage{ifthen}
\usepackage{cite}
\usepackage[cmex10]{amsmath} 
\usepackage{amssymb}

\usepackage{graphicx}

\newtheorem{theorem}{Theorem}

\newtheorem{lemma}{Lemma}

\newcommand{\Pm}{\ensuremath{\mathbb{P}}}
 
\begin{document}
\title{Synergy and Redundancy Dominated Effects in Time Series via Transfer Entropy Decompositions}


 

\author{%
  \IEEEauthorblockN{Jan {\O}stergaard and Payam Shahsavari Boubakani}
  \IEEEauthorblockA{Department of Electronic Systems\\
                    Aalborg University\\
                    Aalborg, Denmark\\
                    Email: \{jo,pasba\}@es.aau.dk}
}


\maketitle

\begin{abstract}
We present a new decomposition of transfer entropy to characterize the degree
of synergy- and redundancy-dominated influence a
time series  has upon the interaction between
other time series.
We prove the existence of a class of time series, where the early past of the conditioning time series yields a synergistic effect upon the interaction, whereas the late past has a redundancy-dominated effect. In general, different parts of the past can have different effects. 
Our information theoretic quantities are easy to compute in practice, and we demonstrate their usage on real-world brain data. 
\end{abstract}

\section{Introduction}

There exists various ways of decomposing differences of conditional mutual informations \cite{Williams:2010,Harder:2013a}. One common decomposition is $\mathrm{II}=I(X;Y)-I(X;Y|Z)$, which is referred to as interaction information \cite{Fano:1961,McGill}.
The interaction information measures the influence the random variable $Z$ has upon the shared information between the random variables $X$ and $Y$ \cite{Ting,yeung:1991}. If $Z$ strengthen the dependency between $X$ and $Y$ then $I(X;Y|Z)>I(X;Y)$. In this case $\mathrm{II}$ is negative and is said to quantify the degree of so-called \emph{synergistic} information within the triplet of variables. On the other hand, if $\mathrm{II}$ is positive, then $Z$ is able to explain some of the dependency between $X$ and $Y$, and $\mathrm{II}$ is then said to quantify the amount of \emph{redundancy}. 

In the neuroscience literature there has been a great interest in obtaining information theoretic decompositions that allows one to easily quantify degrees of synergistic and redundant information exchanges from brain data~\cite{Schneidman,rosas:2019,rosas:2020,mediano:2021,varley:2022,luppi:2022,Scagliarini:2022}.




In this paper, we are interested in characterizing the potential degree of synergy- and redundancy-dominated influence a time-series $\mathcal{S}$ has upon the directional couplings between other interconnected time-series $\mathcal{X}, \mathcal{Y},$ and $\mathcal{Z}$. 
Assume that $\mathcal{S}$ is able to influence $\mathcal{X}, \mathcal{Y},$ and $\mathcal{Z}$. Then, we are interested in understanding whether the change in information flow between  $\mathcal{X}$ and $\mathcal{Y}$, which is due to $\mathcal{S}$ but not due to $\mathcal{Z}$ can be characterized as mainly being synergy- or redundancy-dominated. An application of such a measure could relate to understanding the synergistic effects within the brain. Assume that we are observing different brain regions of a subject, and that some external signal is stimulating the subjects brain. To what degree will such an external stimuli cause a change of synergy- or redundancy-dominated information exchange between the brain regions? Moreover, how do we exclude the effect of other brain regions? We provide an example of this in Section~\ref{sec:examples}.

\section{Notation}
We will frequently refer to the following non-linear  system:
\begin{align}\label{eq:dyn_x}
X_i &= \alpha^{xx}_1 f_{xx}(X_{i-1}) + \alpha^{zx}_2 f_{zx}(Z_{i-2}) + \alpha^{sx}_1 f_{sx}(S_{i-1}) + W^x_i \\  \notag 
Y_i &= \alpha^{yy}_1 f_{yy}(Y_{i-1}) + \alpha^{xy}_2 f_{xy}(X_{i-2}) + \alpha^{zy}_2 f_{zy}(Z_{i-2}) \\ \label{eq:dyn_y}
&\quad + \alpha^{sy}_1 f_{sy}(S_{i-1}) + W^y_i  \\ \label{eq:dyn_z}
Z_i &=  \alpha^{zz}_1 f_{zz}(Z_{i-1}) +  W^z_i \\ \label{eq:dyn_s}
S_i &=  \alpha^{ss}_1 f_{ss}(S_{i-1}) + \alpha^{ss}_2 f_{ss}(S_{i-2})  + W^s_i,
\end{align}
where $f_{\cdot,\cdot}$ are arbitrary (non-linear) differentiable functions with convergent Taylor series, and where $W^\phi_i,W^{\phi'}_{\ell},  \phi\neq\phi' \in \{x,y,z,s\}$ are mutually independent for all $i,\ell$, and otherwise arbitrarily distributed. 
Let $i$ denote the current time instance, and let $X^{i-1} = X_1,\dotsc, X_{i-1}$ denote the sequence of $(i-1)$ past samples of the process. Similarly notation applies to $Y,Z,$ and $S$. 
Let $A$ and $B$ be random variables having elements in the alphabets $\mathcal{A}$ and $\mathcal{B}$, respectively. If $A$ and $B$ are discrete random variables, then we use the notation $\Pm_{A|B}(a|b)$ to denote the probability that $A$ takes on the outcome $a \in \mathcal{A}$ conditioned upon that $B=b\in \mathcal{B}$.

\section{New Synergy and Redundancy Measures}

\subsection{Conditional Mutual Information}
\begin{figure}[th]
    \centering
    \includegraphics[width=5cm]{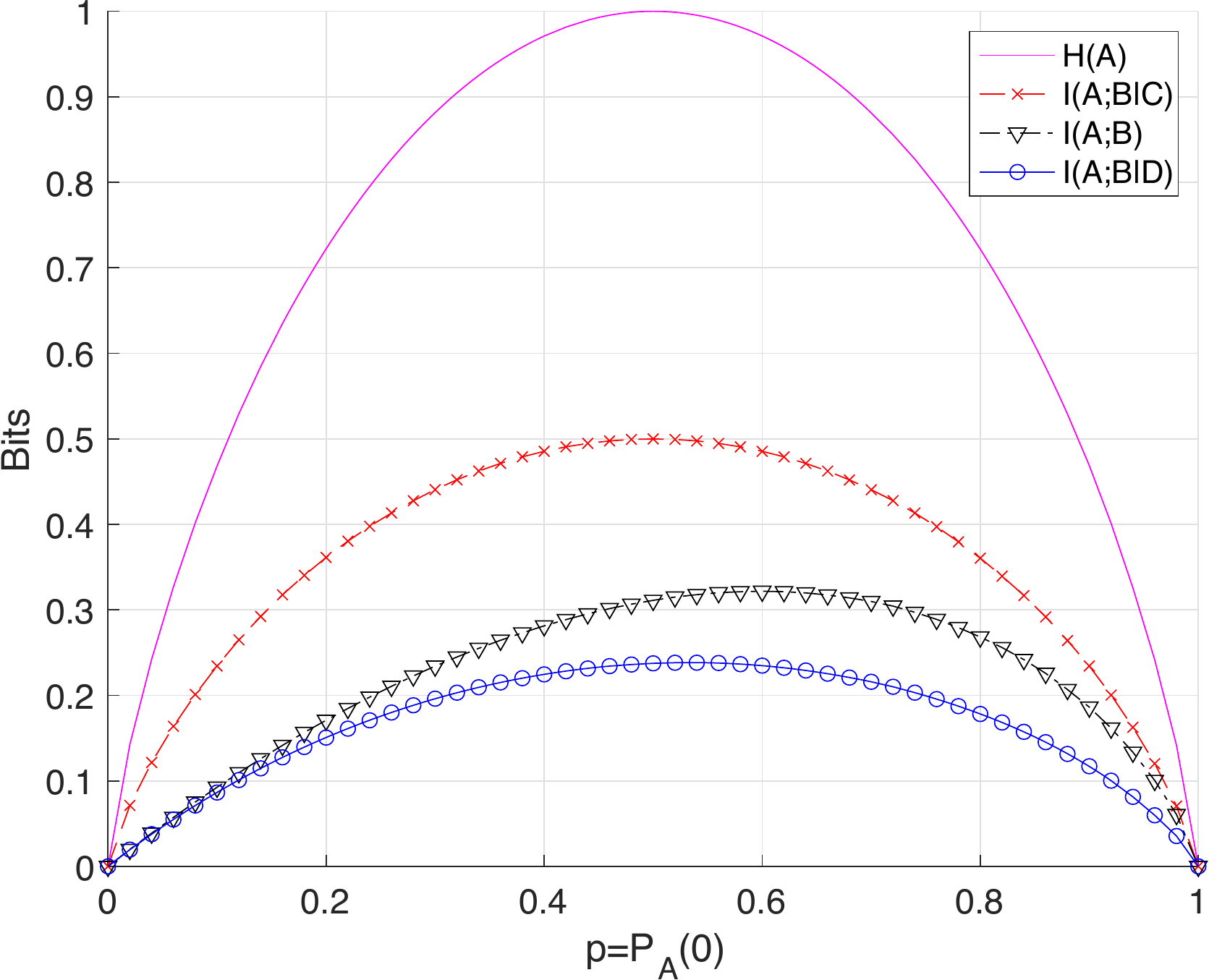}
    \caption{The conditional mutual information exhibits mainly synergistic $I(A;B|C)>I(A;B)$ and redundancy-dominated  $I(A;B|D)<I(A;B)$ effects, when conditioning upon $C$ and $D$, respectively.}
    \label{fig:discrete_PID}
\end{figure}

The mutual information $I(A;B)$ measures the degree of dependency between two random variables $A$ and $B$. If $A$ and $B$ are independent, then $I(A;B)=0$. If they are dependent then $I(A;B)>0$. If we condition upon a third random variable, say $C$,  the resulting conditional mutual information $I(A;B|C)$ reflects  a potential decrease of redundant information as well as a potential increase of synergistic information. 
This is demonstrated below and illustrated in Fig.~\ref{fig:discrete_PID}.
\begin{theorem}
Let $A,C,C'$ be mutually independent binary random variables, where $C,C'$ are uniformly distributed on $\{0,1\}$, and $A$ is arbitrarily distributed with $p=\Pm_A(0)=1-\Pm_A(1)$. 
Let $B=A \wedge C$ and $D = A \wedge C'$, where $\wedge$ denotes the Boolean logical AND operator. Then, for any $ 0\leq p \leq 1$:
\begin{equation}\label{eq:disc}
H(A) \geq I(A;B|C) \overset{(a)}{\geq} I(A;B) \overset{(b)}{\geq} I(A;B|D)
\end{equation}
with equality in $(a)$ and $(b)$ if and only if $p=0$ or $p=1$. 
\end{theorem}

\subsection{Causally Conditioned Transfer Entropy}
The causally conditioned transfer entropy is defined for stationary processes in the following way \cite{schreiber:2000,Kramer:2003}:
\begin{align}
\mathrm{TE}(\mathcal{X} \!\!\to\!\! \mathcal{Y} \| \mathcal{Z}) \triangleq 
I(X^{i-1} ;  Y_i |Y^{i-1}, Z^{i-1}). 
\end{align}
At time $i$, the conditional transfer entropy quantifies the conditional mutual information between the past $X^{i-1}$ and the current $Y_i$ given knowledge of the past $Y^{i-1}$ and $Z^{i-1}$. 

\subsection{Synergistic Versus Common Information}
We  turn our attention to the non-linear continuous-alphabet dynamical system  given in \eqref{eq:dyn_x} -- \eqref{eq:dyn_s}.  Theorem \ref{theo:2} below demonstrates that conditioning upon different parts of the past of a process can have different effects upon the conditional mutual information between other processes. 

\begin{theorem}\label{theo:2}
Consider the non-linear dynamical system in \eqref{eq:dyn_x} -- \eqref{eq:dyn_s}. 
Let $\alpha=\alpha_1^s = \alpha_2^s = \alpha_1^{sy}=\alpha_1^{sx}$ and $\sigma^2_{W^s_{i-j}} = c\alpha^{-2}, \forall j$, so that $\alpha^2 \sigma^2_{W^s_{i-j}} = c$. Then, for any $c>0$:
\begin{align}\notag
&\lim_{\alpha\to 0} 
 \mathrm{TE}(\mathcal{Z}\!\! \to\!\! \mathcal{X} \| \mathcal{Y},S_{i-1}) 
>
\lim_{\alpha\to 0} 
 \mathrm{TE}(\mathcal{Z}\!\! \to \!\!\mathcal{X} \| \mathcal{Y})  \\ \label{eq:cond_mi_left}
 &\quad
 > \lim_{\alpha\to 0} 
  \mathrm{TE}(\mathcal{Z}\!\! \to\!\! \mathcal{X} \| \mathcal{Y},S^{i-2}) .
\end{align}
\end{theorem}

\begin{IEEEproof}
The second inequality is proved by Lemma~\ref{lem:syn_right} in the appendix. Thus, we only need to prove that the first inequality holds. To do this we expand the first conditional transfer entropy  in \eqref{eq:cond_mi_left} as follows:
{\allowdisplaybreaks
\begin{align}\notag
&  I(Z^{i-1} ;  X_i |X^{i-1},Y^{i-1},S_{i-1}) \\ \notag
&=    I(Z^{i-1} ;  \alpha^{xx}_1 f_{xx}(X_{i-1}) + \alpha^{zx}_2 f_{zx}(Z_{i-2}) + \alpha^{sx}_1 f_{sx}(S_{i-1}) \\
&\quad + W^x_i  |X^{i-1},Y^{i-1},S_{i-1})  \\ \notag
&=    I(Z^{i-1} ;  \alpha^{zx}_2 f_{zz}(Z_{i-2})  + W^x_i  |X^{i-1},Y^{i-1},S_{i-1})  \\ \notag
&=    I(Z_{i-2} ;  \alpha^{zx}_2 f_{zz}(Z_{i-2})  + W^x_i  |X^{i-1},Y^{i-1},S_{i-1})  \\
&\quad+ I( Z_{i-1},Z^{i-3} ;  W^x_i  |X^{i-1},Y^{i-1},S_{i-1}, Z_{i-2}) \\ \label{eq:l1}
&=    I(Z_{i-2} ;  \alpha^{zx}_2 f_{zz}(Z_{i-2})  + W^x_i  |X^{i-1},Y^{i-1},S_{i-1}). 
\end{align}
}

When we are not conditioning upon $S_{i-1}$, we obtain from \eqref{eq:tmp2l} in the appendix:
\begin{align}\notag
 &I(Z^{i-1} ; X_i|X^{i-1},Y^{i-1}) = I(Z_{i-2} ;  \alpha^{zx}_2 f_z(Z_{i-2})  +  \alpha \beta W_{i-1}^s \\ \label{eq:l2t}
 &\quad + W^x_i  
 + o(\alpha)  |X^{i-1},Y^{i-1}),
\end{align}
where $o(\alpha)/\alpha \to 0$ as $\alpha\to 0$. 
Since $X^{i-1},Y^{i-1}, S_{i-1}$ are independent of $W^x_i$ and $\alpha^2 \sigma_{W^s}^2 = c>0$ it is clear that 
$\alpha^{zx}_2 f_z(Z_{i-2})  +  \alpha \beta W_{i-1}^s + W^x_i$ is a more noisy version of $Z_{i-2}$ than $\alpha^{zx}_2 f_z(Z_{i-2}) + W^x_i$. Thus, \eqref{eq:l1} is strictly smaller than \eqref{eq:l2t}. This proves the theorem.
\end{IEEEproof}


\subsection{Synergy- and Redundancy-Dominated Measures}
We are now in a position to define a notion of maximal coupling strength for the synergistic information occuring between time series due the external stimuli. 
Since conditioning upon different parts of the past lead to different effects upon the conditional transfer entropy, we introduce a maximization: 
\begin{align}\notag
I^{\mathrm{syn}}_S ( \mathcal{X}\!\!\to\!\! \mathcal{Y}\|\mathcal{Z}) &\triangleq 
\max_{f(\hat{S}|S)}  
I(X^{i-1};Y_i | Y^{i-1}, Z^{i-1}, \hat{S})  \\  \label{eq:Isyn}
&\quad -  
I(X^{i-1};Y_i | Y^{i-1}, Z^{i-1}).
\end{align}
The maximization is over all distributions on $\mathcal{S}$ as suggested in \cite{james:2021}. It can be observed that $I^{\mathrm{syn}}_S$ is non-negative, and a positive number indicates synergistic information. 
The reasons for forming a difference of information quantities are twofold. First, the excess synergistic information due to conditioning only upon $\mathcal{Z}$ is in \eqref{eq:Isyn} removed by forming a difference of causally conditioned mutual informations. Thus, we are able to remove the effect that $\mathcal{Z}$ has by itself upon the mutual information between $\mathcal{X}$ and $\mathcal{Y}$. Yet, any potential synergistic information due to the combined knowledge of $\mathcal{S}$ and $\mathcal{Z}$ are not excluded.
Second, it allows us to separate the part of the external stimuli that mainly has a synergistic effect. 
Whilst $\mathcal{S}$ does not explicitly appear in the first term in \eqref{eq:Isyn}, it should be noted that $\mathcal{X}, \mathcal{Y},$ and $\mathcal{Z}$ could potentially be causal functions of $\mathcal{S}$. 
It follows that \eqref{eq:Isyn}  quantifies the part of the information that is synergistically occuring between  $\mathcal{X}$ and $\mathcal{Y}$ due to $\mathcal{S}$ but not due to $\mathcal{Z}$ by itself.


The maximization in \eqref{eq:Isyn} is generally computationally intractable. Instead we introduce a lower bound to \eqref{eq:Isyn}, which is simple to compute on  time-series data as we demonstrate in Section~\ref{sec:examples}. In particular, instead of extremizing over all measurable functions, we simply extremize over subsets of the elements of $\mathcal{S}$ as suggested in \cite{baboukani:2022}. Specifically, we introduce the following information theoretic quantity:
\begin{align}\notag
 & \hat{I}^{\mathrm{syn}}_\mathcal{S} ( \mathcal{X}\!\!\to\!\! \mathcal{Y}\|\mathcal{Z}) \triangleq 
\max_{\hat{\mathcal{S}} \subseteq \mathcal{S} }
\mathrm{TE}(\mathcal{X}\!\!\to\!\! \mathcal{Y} \| \mathcal{Z}, \hat{\mathcal{S}})   -\mathrm{TE}(\mathcal{X}\!\!\to\!\!\mathcal{Y} \| \mathcal{Z})   
\\ \label{eq:isyn_hat}
&=\!\!\!\max_{\hat{\mathcal{S}} \subseteq S^{i-1}} \!\!\!
I(X^{i-1};Y_i | Y^{i-1}\!, Z^{i-1}\!\!, \hat{S})    -
I(X^{i-1};Y_i | Y^{i-1}\!\!, Z^{i-1}). 
\end{align}
Clearly,  $\hat{I}^{\mathrm{syn}}_\mathcal{S} ( \mathcal{X}\!\!\to\!\! \mathcal{Y}\|\mathcal{Z}) \leq I^{\mathrm{syn}}_\mathcal{S} ( \mathcal{X}\!\!\to\!\! \mathcal{Y}\|\mathcal{Z})$, since we restrict the maximization to a subset of the feasible set. 

Efficient mutual information estimators based on variants of the KSG estimator \cite{Kraskov:2004} are known to be biased \cite{Holmes:2019}. This bias depends on the dimensionality and partly on the distributions of the variables \cite{Holmes:2019}. To reduce the effect of bias, we suggest to condition the second term in \eqref{eq:isyn_hat} on the very late past of $S$, in order to guarantee that the two terms are of equal dimensionality and approximately equal distributions, that is:
\begin{align*}
\!\!\!\max_{\hat{\mathcal{S}} \subseteq S^{i-1}} \!\!\!
I(X^{i-1};Y_i | Y^{i-1}\!, Z^{i-1}\!, \hat{S} )   
-   
I(X^{i-1};Y_i | Y^{i-1}\!, Z^{i-1}\!,\tilde{S}), 
\end{align*}
where $\tilde{S} = S_{i-T+1}, \dotsc, S_{i-T+|\hat{S}|}$, denotes $|\hat{S}|$ samples from the very late past $T\gg 1$ of $S$. Thus, $\tilde{S}$ has the same dimensionality as $\hat{S}$ and approximately the same distribution. For $T$ sufficiently large, the statistical dependency between $\tilde{S}$ and $X,Y,Z$, is minimal, and therefore conditioning upon $\tilde{S}$ will only have a minor effect upon the true transfer entropy.

To characterize the maximal redundancy-dominated effect, we introduce a minimization over $\mathcal{S}$ and form the following non-negative quantity and a corresponding lower bound:
\begin{align*}\notag
&I^{\mathrm{red}}_S(\mathcal{X}\!\!\to\!\! \mathcal{Y}\|\mathcal{Z}) \triangleq 
\mathrm{TE}(\mathcal{X}\!\!\to\!\! \mathcal{Y} \| \mathcal{Z})  
\!-\! \!  \min_{f(\hat{S}|S)} \!\! \mathrm{TE}(\mathcal{X}\!\!\to\!\! \mathcal{Y} \| \mathcal{Z},\hat{\mathcal{S}} ) \\ \notag
&\geq 
\hat{I}^{\mathrm{red}}_S(X\!\!\to\!\! Y\|Z) 
\triangleq 
\mathrm{TE}(\mathcal{X}\!\!\to\!\! \mathcal{Y} \| \mathcal{Z})  
-  \min_{\hat{\mathcal{S}} \subseteq \mathcal{S}}\mathrm{TE}(\mathcal{X}\!\!\to\!\! \mathcal{Y} \| \mathcal{Z},\hat{\mathcal{S}} ) \\ 
&=
 I(X^{i-1};Y_i | Y^{i-1}\!, Z^{i-1}) - \!\!\!\min_{\hat{\mathcal{S}} \subseteq S^{i-1}} \!\! I(X^{i-1};Y_i | Y^{i-1}\!\!, Z^{i-1}\!\!, \hat{S}),
\end{align*}
where one can also decide to condition upon $\tilde{S}$ in the first term to reduce the effect of estimator bias. 
Depending upon the sign of the difference $\hat{I}^{\mathrm{red}}_\mathcal{S} - \hat{I}^{\mathrm{syn}}_\mathcal{S}$, the overall effect of $\mathcal{S}$ is mainly synergy or redundancy-dominated.


\section{Example on Brain Data}\label{sec:examples}

Locating areas in the brain that are causing seizures is an important issue in the study of epilepsy in humans.
We consider intracranial EEG recordings from a patient with drug-resistant epilepsy.
The data were recorded by an implanted array of $8\times 8$ cortical electrodes referred to as  $E(1), \dotsc, E(64)$, and two left hippocampal depth electrodes (each having six electrodes), referred to as $E(65), \dotsc, E(76)$.
The time-series data from electrode $i$ are denoted $E(i) = E_1(i), \dotsc, E_N(i)$.
The data are available from \cite{earth:} and described in details in \cite{kramer:2008}.
A total of eight epileptic seizures (Ictal) and eight pre-ictal periods were recorded. The EEG data were downsampled to 100 Hz. In the examples, we used the TRENTTOOL conditional mutual information estimator with $k=10$ nearest neighbors \cite{lindner:2011}.


\subsection{Effect of Early Versus Late Past}
\begin{figure}[th]
    \centering
    \includegraphics[width=4.5cm]{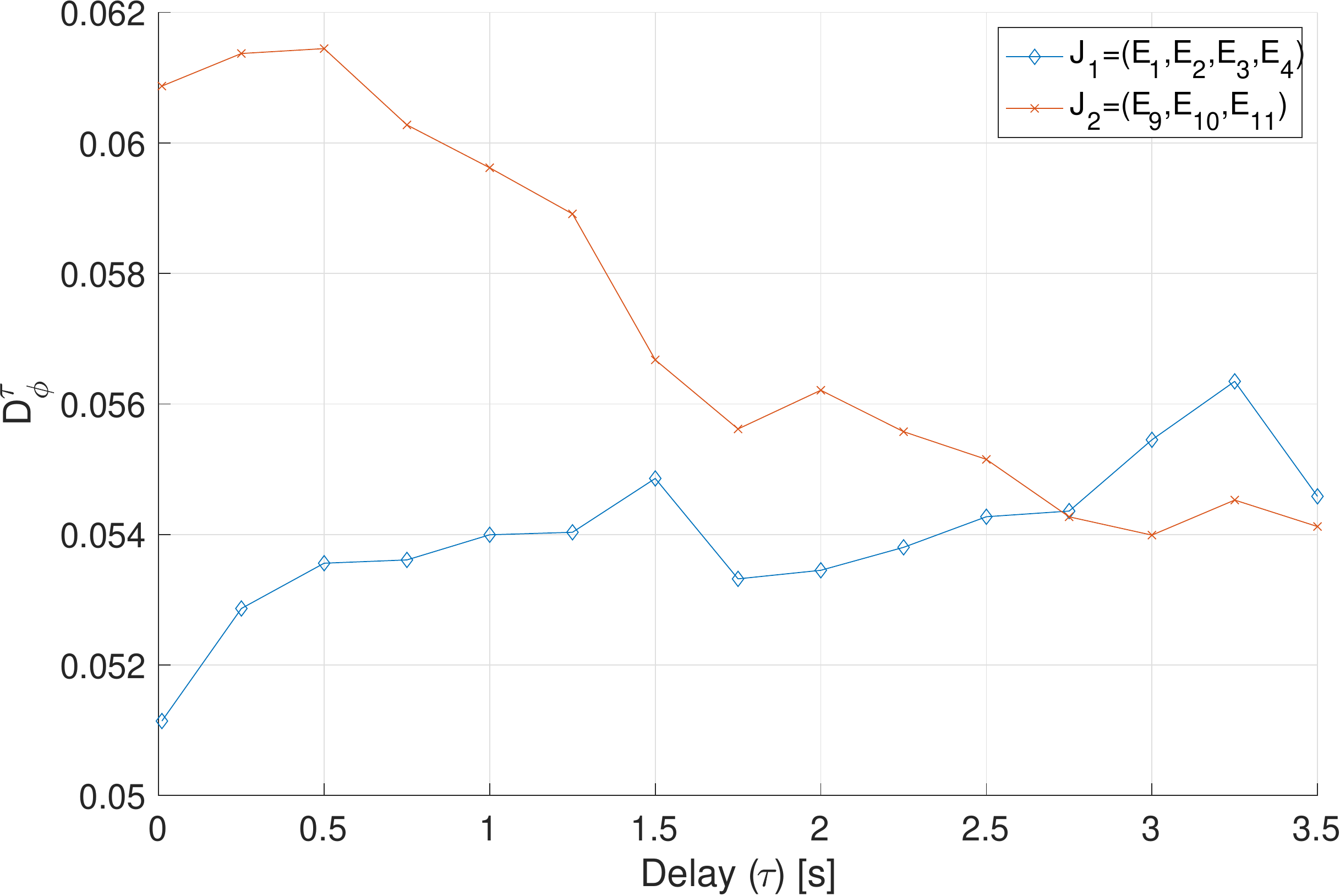}
    \includegraphics[width=4cm]{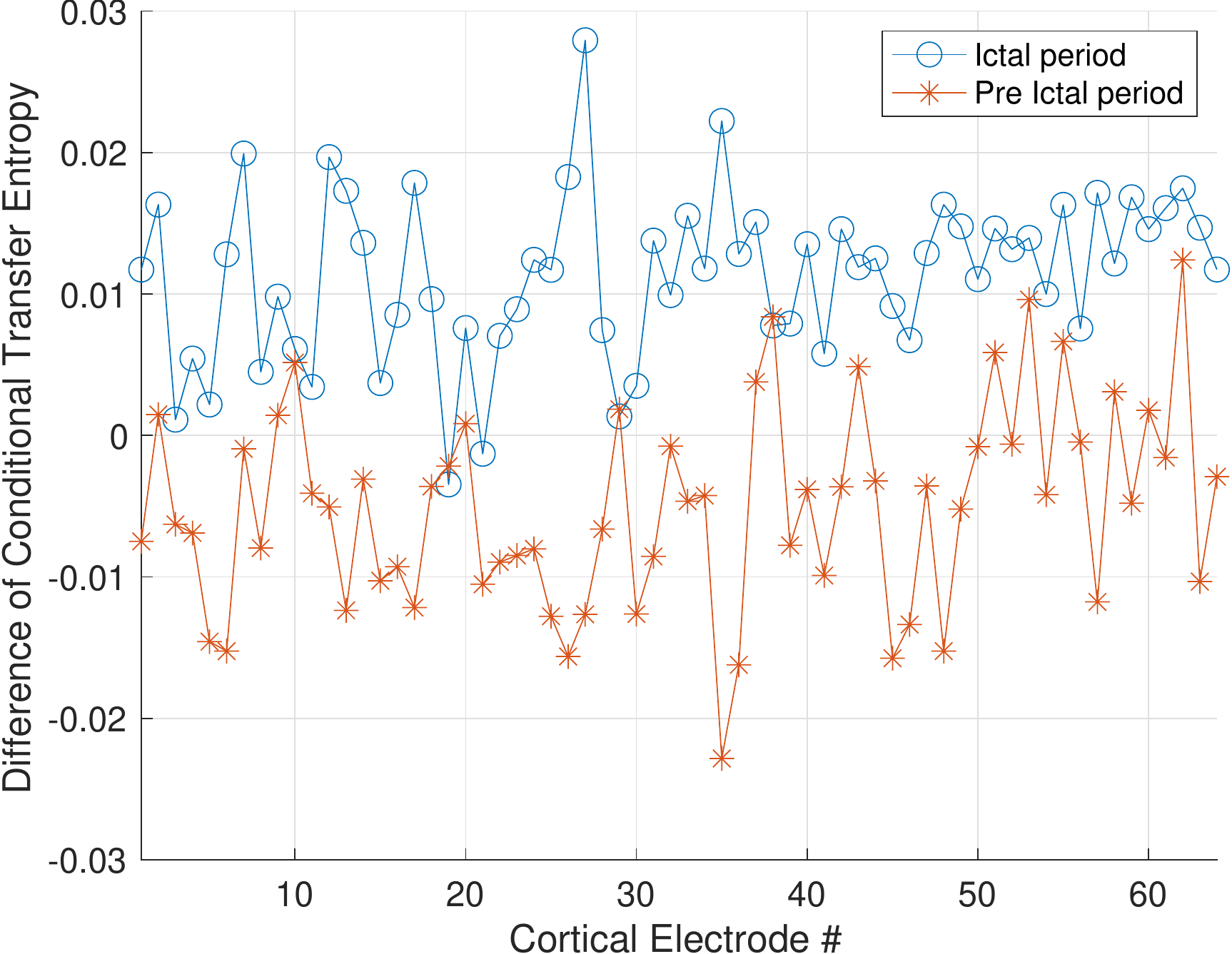}
    \caption{Left: $D^\tau_\phi$ in \eqref{eq:Dtau} as a function of delay $\tau$ and subset $\mathcal{J}_\phi, \phi=1,2$. Right: The difference $\hat{I}_{E(76)}^{\mathrm{syn}}(i) - \hat{I}_{E(75)}^{\mathrm{syn}}(i)$ for $i=1,\dotsc, 64$. Positive numbers indicate that the effect of electrode $E(76)$ dominates over that of $E(75)$, whereas the opposite is true for negative numbers. }
    \label{fig:ictal}
\end{figure}

In \cite{Stramaglia:2014}, it was observed that the depth electrodes and the lower left corner of the cortical grid with electrodes 1 -- 4, 9 -- 11, and 17 exhibited strong synchronous neuronal activity during seizures. In our results illustrated in Fig.~\ref{fig:ictal}, we focus on the two subsets $\mathcal{J}_1= \{1,2,3,4\}$ and $\mathcal{J}_2 = \{9,10,11\}$ of cortical electrodes. 
Let
$E_{\tau}(\ell)$ be the $\tau$th delayed sample of the $\ell$th electrode, and let us compute:
\begin{align}\label{eq:Dtau}
   D^{\tau}_\phi \triangleq \frac{1}{|\mathcal{L}|}\sum_{\ell\in \mathcal{L}}\ \sum_{\substack{i,j \in \mathcal{J}_\phi\\ i\neq j }}\mathrm{TE}(E(i)\!\! \to\!\! E(j) \| E_{\tau}(\ell)),
\end{align}
where $\mathcal{L}=\{65, \dotsc, 76\}$ denotes all depth electrodes, and where $i,j$ are both in $\mathcal{J}_1$ or both in  $\mathcal{J}_2$. 
In Fig.~\ref{fig:ictal}, the effect of $\tau$ and $\phi$ on $D^\tau_\phi$ can be observed. Interestingly, conditioning upon the early past (before 0.5 s) of $\mathcal{J}_1$ or $\mathcal{J}_2$, results in a redundancy- or synergy-dominated effect, respectively. 
At larger delays, $D^\tau_1$ and $D^\tau_2$ get closer to each other.

\subsection{Greatest Synergistic Effect}
In \cite{Stramaglia:2014}, it was suggested that synergistic interaction between depths and cortical electrodes were greatest from depth electrodes $E(75)$ and $E(76)$; the last two electrodes in the second depth electrode. In this example, we identify whether $E(75)$ or $E(76)$ causes the greatest synergistic effect as quantified by $\hat{I}^{\mathrm{syn}}$ in \eqref{eq:isyn_hat}. 
We used \eqref{eq:isyn_hat} to estimate the synergistic information $\hat{I}_{E(76)}^{\mathrm{syn}}(i) \triangleq \hat{I}_{E(76)}^{\mathrm{syn}} (\mathcal{X}\!\!\to\!\! \mathcal{Y}\| E(75) )$, which is due to $E(76)$ and not due to $E(75)$. Here $\mathcal{X}=E(i), i=1,\dotsc, 64$, and $\mathcal{Y} = \{E(1),E(2),\cdots ,E(64)\}\backslash\{\mathcal{X}\}$. Thus, we measured the synergistic information from the $i$th electrode to the remaining 63 cortical electrodes. We similarly estimated $\hat{I}_{E(75)}^{\mathrm{syn}}(i) \triangleq \hat{I}_{E(75)}^{\mathrm{syn}} (\mathcal{X}\!\!\to\!\! \mathcal{Y}\| E(76) )$. Their difference $\hat{I}_{E(76)}^{\mathrm{syn}}(i) - \hat{I}_{E(75)}^{\mathrm{syn}}(i)$ for $i=1,\dotsc, 64,$ is plotted in Fig.~\ref{fig:ictal} for the Ictal (circles) and pre-Ictal (stars) periods. During seizures (Ictal period), it can be observed that the effect of $E(76)$  dominates that of $E(75)$ in the sense that the difference $\hat{I}_{E(76)}^{\mathrm{syn}}(i) - \hat{I}_{E(75)}^{\mathrm{syn}}(i)$ is positive for most electrodes, and the average difference is 0.011, which is positive. 
This suggests a greater emergence of synergistic information exchange between cortical brain regions due to stimulation by $E(76)$ than  $E(75)$ during seizures. 



\section{Conclusions}
We presented a new decomposition of transfer entropy, which made it possible to quantify the effect different parts of the past of a time series has upon the transfer entropy between other time series. 
We applied the decomposition to intracranial EEG recordings, and observed that the early past can have a synergistic effect. This is interesting, since generally the early past carry redundant information due to volume conduction effects, i.e., where each EEG channel records the instantaneous linear mixing of multiple brain source activities \cite{He:2011}.

\appendix

\begin{IEEEproof}[Proof of Theorem 1]
We first consider the left hand side of \eqref{eq:disc} and rewrite $I(A;B|C) = H(A|C) - H(A|B,C) = H(A) - H(A|B,C)$, where the latter conditional entropy can be further re-written as:
{\allowdisplaybreaks
\begin{align*}
&H(A|B,C) = \sum_{b,c} \Pm_{B,C}(b,c) H(A|B=b,C=c) \\
&= -\Pm_{B,C}(0,0)\big(\Pm_{A|B,C}(0|0,0)\log_2( \Pm_{A|B,C}(0|0,0))    \\
&\quad +\Pm_{A|B,C}(1|0,0)\log_2( \Pm_{A|B,C}(1|0,0) ) \big)    \\
&\quad -\Pm_{B,C}(0,1)\big(\Pm_{A|B,C}(0|0,1)\log_2( \Pm_{A|B,C}(0|0,1))    \\
&\quad +\Pm_{A|B,C}(1|0,1)\log_2( \Pm_{A|B,C}(1|0,1) ) \big)    \\
&\quad -\Pm_{B,C}(1,0)\big(\Pm_{A|B,C}(0|1,0)\log_2( \Pm_{A|B,C}(0|1,0))    \\
&\quad +\Pm_{A|B,C}(1|1,0)\log_2( \Pm_{A|B,C}(1|1,0) ) \big)    \\
&\quad -\Pm_{B,C}(1,1)\big(\Pm_{A|B,C}(0|1,1)\log_2( \Pm_{A|B,C}(0|1,1))    \\
&\quad +\Pm_{A|B,C}(1|1,1)\log_2( \Pm_{A|B,C}(1|1,1) ) \big)    \\
&= -\Pm_{B,C}(0,0)\big(\Pm_{A|B,C}(0|0,0)\log_2( \Pm_{A|B,C}(0|0,0))    \\
&\quad +\Pm_{A|B,C}(1|0,0)\log_2( \Pm_{A|B,C}(1|0,0) ) \big)    \\
&\quad -\Pm_{B,C}(0,1)\big(\Pm_{A|B,C}(0|0,1)\log_2( \Pm_{A|B,C}(0|0,1))    \\
&\quad +\Pm_{A|B,C}(1|0,1)\log_2( \Pm_{A|B,C}(1|0,1) ) \big)    \\
&= -\frac{1}{2}\big( p\log_2(p) + (1-p)\log_2(1-p)   \big)   \\
&=\Pm_{B,C}(0,0)H(A) = 0.5 H(A),
\end{align*}}
which follows since $\Pm_{B,C}(0,0) = p, \Pm_{B,C}(0,1) = \frac{1}{2}(1-p), \Pm_{B,C}(1,0)=0, 
\Pm_{A|B,C}(0|1,0)=0, \Pm_{A|B,C}(1|1,1)=1, \Pm_{A|B,C}(0|0,1)=p$, and $\Pm_{A|B,C}(1|0,1)=1-p.$ 
It follows:
\begin{align}
    I(A;B|C) = 0.5H(A). 
\end{align}

We now consider the middle term in \eqref{eq:disc}, i.e., $I(A;B) = H(A)-H(A|B)$, where the latter term can be written as:
{\allowdisplaybreaks
\begin{align*} \notag
&H(A|B) \overset{(a)}{=} \sum_{b\in \{0,1\}} \Pm_B(b) H(A|B=b) \\
&= - \sum_{b\in \{0,1\}} \sum_{a\in \{0,1\}}\Pm_B(b) \Pm_{A|B}(a|b)\log_2(\Pm_{A|B}(a|b))  \\
&=-\Pm_B(0) \Pm_{A|B}(0|0)\log_2( \Pm_{A|B}(0|0) ) \\
&\quad -\Pm_B(0)  \Pm_{A|B}(1|0)\log_2( \Pm_{A|B}(1|0) )  \\
&\quad -\Pm_B(1)  \Pm_{A|B}(0|1)\log_2( \Pm_{A|B}(0|1) ) \\
&=-\Pm_A(0)\log_2\frac{\Pm_A(0)}{\Pm_B(0)} + \frac{1}{2} \Pm_A(1)\log_2\frac{1}{2}\frac{\Pm_A(1)}{\Pm_B(0)} \\
&\quad - \frac{1}{2}\Pm_A(1)\log_2 \frac{1}{2}\frac{\Pm_A(1)}{\Pm_B(1)} \\
&=H(A) + \Pm_A(0)\log_2(\Pm_B(0)) \\
&\quad + \frac{1}{2}\Pm_A(1)(\log_2(\Pm_B(0)) + \log_2(\Pm_B(1)) + \Pm_A(1) \\
&= \frac{1}{2}H(A) + \varphi
\end{align*}}
where
\begin{align*}
\varphi &= \frac{1}{2}H(A) + \Pm_A(0)\log_2(\Pm_B(0)) \\
&\quad + \frac{1}{2}\Pm_A(1)(\log_2(\Pm_B(0)) + \log_2(\Pm_B(1)) + \Pm_A(1).
\end{align*}
The second derivative of $\varphi$ with respect to $p=\Pm_B(0)$ is:
\begin{align}\label{eq:phi}
    \frac{\partial^2}{\partial p^2} \varphi = - \frac{1}{2(1-p)p\ln(2)}. 
\end{align}
It follows that $\varphi$ is a strictly concave function in $p$ and to find its minima it is sufficient to check the two boundary points $p=0$ and $p=1$, which when inserted into \eqref{eq:phi} yields $\varphi_{|p=0} = \varphi_{|p=1} =0$. Thus, $\varphi$ is a non-negative function. We can now show that  $I(A;B)<I(A;B|C)$:
\begin{align}
I(A;B) &= H(A) - H(A|B) 
= H(A) - \frac{1}{2}H(A) - \phi \\
&\overset{(a)}{\leq} \frac{1}{2}H(A) = I(A;B|C),
\end{align}
with equality in $(a)$ only for $p =0$ or $p=1$, which follows since the gradient of $\varphi$ is non-zero at these extremes. 

Let us now consider the last term in \eqref{eq:disc}:
\begin{align*}
    &I(A;B|D) = H(A|D) - H(A|B,D) \\
    &= \sum_{d} \Pm_{D}(d) H(A|D\!=\!d) \!- \!\sum_{b,d} \Pm_{B,D}(b,d) H(A|B\!=\!b,D\!=\!d).
\end{align*}

Since we are considering binary alphabets, we can express
the probabilities explicitly and determine, which ones that are trivially 0 or 1 -- in which case they can be excluded.
{\allowdisplaybreaks
\begin{align*}
&\Pm_{A|D}(0|0) = \Pm_{D|A}(0|0)p/\Pm_{D}(0) = p/\Pm_{D}(0). \\
&\Pm_{A|D}(1|0) = \frac{1}{2}(1-p)/\Pm_{D}(0), P_{B,D}(0,0) = \frac{1}{4}(1-p) + p. \\
&P_{B,D}(0,1) =  \frac{1}{4}(1-p), \Pm_{A|B,D}(0|0,0) = \frac{p}{\Pm_{B,D}(0,0)}. \\
&\Pm_{A|B,D}(0|1,0) = 0, \Pm_{A|B,D}(1|0,0) = \frac{1}{4}\frac{(1-p)}{\Pm_{B,D}(0,0)}.\\
&\Pm_{A|B,D}(1|1,0) = 1.
\end{align*}}

We can now expand $H(A|B,D)$ as a function of $p$:
{\allowdisplaybreaks
\begin{align*}
    &H(A|B,D) = -\Pm_{B,D}(0,0)\\
    &\quad\times\bigg( \frac{p}{\Pm_{B,D}(0,0)} \log_2\frac{p}{\Pm_{B,D}(0,0)} \\
    &\quad+ \frac{1}{4}\frac{(1-p)}{\Pm_{B,D}(0,0)} \log_2\frac{1}{4}\frac{1-p}{\Pm_{B,D}(0,0)}  \bigg)\\
    &= -\bigg( p \log_2\frac{p}{\Pm_{B,D}(0,0)} 
  + \frac{1}{4}(1-p) \log_2\frac{1}{4}\frac{(1-p)}{\Pm_{B,D}(0,0)}  \bigg)\\
  &= - p \log_2\frac{p}{p +\frac{1}{4}(1-p)}  - \frac{1}{4}(1-p) \log_2\frac{1}{4}\frac{(1-p)}{p+\frac{1}{4}(1-p)} \\
   &= \frac{1}{2}(1-p) + (p+\frac{1}{4}(1-p))\log_2( p+\frac{1}{4}(1-p)) \\
   &\quad - \frac{3}{4}p \log_2(p) + \frac{1}{4}H(A).
\end{align*}}

Similarly, we can express $H(A|B)$ as a function of $p$:
\begin{align}
    &H(A|D) = -\Pm_{D}(0)\big( \Pm_{A|D}(0|0)\log_2(  \Pm_{A|D}(0|0) ) \\
    &\quad +  \Pm_{A|D}(1|0)\log_2(  P_{A|D}(1|0) \big) \\
    &= - p\log_2  \frac{p}{\Pm_{D}(0)} - \frac{1}{2}(1-p)\log_2  \frac{\frac{1}{2}(1-p)}{\Pm_{D}(0)},
\end{align}
where $\Pm_{D}(0) = 1/2+p/2$. 
We have now expressed $I(A;B)$ and $I(A;B|D)$ as simple functions of $p$.  This allows us to easily determine convexity by use of the second derivative of $I(A;B) - I(A;B|D)$ with respect to $p$, that is:
\begin{align}
\frac{\partial^2}{\partial p^2} (I(A;B) - I(A;B|D)) = \frac{1}{(3p^3 - 2p^2-p)\ln(2)}.
\end{align}
It is clear that $\frac{\partial^2}{\partial p^2}(3p^3 - 2p^2-p)= 18p$, which implies that $3p^3 - 2p^2-p$ is strictly convex. Thus, the maxima are at the boundaries, and these are furthermore the minima of $I(A;B) - I(A;B|D)$. Inserting $p=0$ and $p=1$ into $I(A;B)-I(A;B|D)$ yields $0$, which shows that $I(A;B)-I(A;B|D)$ is a non-negative function, and it is zero only at the boundaries. 
This proves the theorem. 
\end{IEEEproof}

\begin{lemma}\label{lem:syn_right}
Consider \eqref{eq:dyn_x} -- \eqref{eq:dyn_s}. 
Let $\alpha=\alpha_1^s = \alpha_2^s = \alpha_1^{sy}=\alpha_1^{sx}$ and $\sigma^2_{W^s_{i-j}} = c\alpha^{-2}, \forall j$, so  $\alpha^2 \sigma^2_{W^s_{i-j}} = c$. Then, for $c>0$:
\begin{align*}\label{eq:cond_mi_right}
\lim_{\alpha\to 0} 
 &I(Z^{i-1} ;\!  X_i |X^{i-1}\!\!,Y^{i-1}) 
 \! > \!\lim_{\alpha\to 0} \!
 I(Z^{i-1} ;\!  X_i |X^{i-1}\!\!,Y^{i-1}\!\!,S^{i-2}).
\end{align*}
\end{lemma}
\begin{IEEEproof}
    Let us first consider the left term:
\begin{align*}\notag
&I(Z^{i-1} ;  X_i |X^{i-1},Y^{i-1}) =    I(Z^{i-1} ;  \alpha^{xx}_1 f_x(X_{i-1}) \\ \notag
&+ \alpha^{zx}_2 f_z(Z_{i-2}) + \alpha^{sx}_1 f_s(S_{i-1}) + W^x_i  |X^{i-1},Y^{i-1})  \\ \notag
&=    I(Z^{i-1} ;  \alpha^{zx}_2 f_z(Z_{i-2})  + \alpha^{sx}_1 f_s(S_{i-1}) + W^x_i  |X^{i-1},Y^{i-1})  \\ \notag
&\overset{(a)}{=}    I(Z_{i-2} ;  \alpha^{zx}_2 f_z(Z_{i-2})  + \alpha^{sx}_1 f_s(S_{i-1}) + W^x_i  |X^{i-1},Y^{i-1}) \\ \notag
&\quad+ I( Z_{i-1},Z^{i-3} ;    \alpha^{sx}_1 f_s(S_{i-1}) + W^x_i  |X^{i-1},Y^{i-1}, Z_{i-2}) \\ 
&\overset{(b)}{=}    I(Z_{i-2} ;  \alpha^{zx}_2 f_z(Z_{i-2})  + \alpha^{sx}_1 f_s(S_{i-1}) + W^x_i  |X^{i-1},Y^{i-1}), 
\end{align*}
where $(a)$ follows from the chain rule of mutual information and $(b)$ follows since $W_i^x, S_{i-1}, Z^{i-1}$ are mutually independent and also conditionally independent given $X^{i-1},Y^{i-1},Z_{i-2}$.  
At this point, let $\alpha=\alpha_1^s = \alpha_2^s = \alpha_1^{sy}=\alpha_1^{sx}$ and $\sigma^2_{W^s_{i-j}} = c\alpha^{-2}, \forall j$, so that $\alpha^2 \sigma^2_{W^s_{i-j}} = c$. The first-order (linear) series expansion of $f_s$ is: $f_s(S_{i-1}) = \beta W_{i-1}^s + o(\alpha)$, for some constant $\beta$. We continue as follows:
\begin{align} \notag
&I(Z_{i-2} ;  \alpha^{zx}_2 f_z(Z_{i-2})  +  \alpha f_s(S_{i-1})  + W^x_i   |X^{i-1},Y^{i-1}) \\ 
&= 
I(Z_{i-2} ;  \alpha^{zx}_2 f_z(Z_{i-2})  +  \alpha \beta W_{i-1}^s + W^x_i  \\ \label{eq:tmp2l}
&\quad + o(\alpha)  |X^{i-1},Y^{i-1}) \\
&= I(Z_{i-2} ;  \alpha^{zx}_2 f_z(Z_{i-2})  +  V + o(\alpha)  |F),
\end{align}
where $F=\{X^{i-1}, Y^{i-1}\}$ and $V=\alpha \beta W_{i-1}^s + W^x_i$. 
Note that $F$ is independent of $V$ and $F$ is asymptotically independent of the terms within $o(\alpha)$ in the limit as $\alpha\to 0$. It follows that for small $\alpha$, conditioning upon $F$ results in very little synergistic information, since the noise $V$ added to $f_z(Z_{i-2})$ cannot be predicted (and hence reduced) from knowledge of $F$. 

Consider now the right side of the inequality in the lemma:
{\allowdisplaybreaks
\begin{align} \notag 
&I(Z^{i-1} ;  X_i |X^{i-1},Y^{i-1},S^{i-2}) \\ \notag
&=    I(Z^{i-1} ;  \alpha^{xx}_1 f_x(X_{i-1}) + \alpha^{zx}_2 f_z(Z_{i-2}) + \alpha^{sx}_1 f_s(S_{i-1}) \\ \notag 
&\quad + W^x_i  |X^{i-1},Y^{i-1},S^{i-2})  \\ \notag 
&=    I(Z^{i-1} ;  \alpha^{zx}_2 f_z(Z_{i-2})  + \alpha^{sx}_1 \beta W_{i-1}^s + W^x_i \\ \notag
&\quad+ o(\alpha) |X^{i-1},Y^{i-1},S^{i-2})  \\ \notag 
&=    I(Z_{i-2} ;  \alpha^{zx}_2 f_z(Z_{i-2})  +  \alpha^{sx}_1 \beta W_{i-1}^s + W^x_i \\ \notag
&\quad + o(\alpha) |X^{i-1},Y^{i-1},S^{i-2}) \\ \notag 
&\quad+ I( Z_{i-1},Z^{i-3} ;    \alpha^{sx}_1 \beta W_{i-1}^s + W^x_i \\ \notag 
&\quad + o(\alpha) |X^{i-1},Y^{i-1},S^{i-2},Z_{i-2}) \\ 
&=    I(Z_{i-2} ;  \alpha^{zx}_2 f_z(Z_{i-2}) + \alpha^{sx}_1 \beta W_{i-1}^s \\ \notag
&\quad + W^x_i + o(\alpha) |X^{i-1},Y^{i-1},S^{i-2}) \\ \notag 
&= I(Z_{i-2} ;  \alpha^{zx}_2 f_z(Z_{i-2})  +  V  + o(\alpha) |F,S^{i-2}).
\end{align}}
Since $S^{i-2}$ and $F$ are independent of $W_{i-1}^s$, it follows that knowing $S^{i-2}$ and $F$ do not provide further synergistic information than what is possible knowing only $F$.

Let us now consider the potential additional reduction in common information between $Z_{i-2}$ and $\alpha^{zx}_2 f_z(Z_{i-2})  +  V$ due to knowledge of $S^{i-2}$ in addition to $F$. We define $\phi$ as a conditional sufficient statistics of $X_{i-1}$ and $Y_{i-1}$ with respect to $Z_{i-2}$ conditioned upon $F'=F\backslash\{X_{i-1},Y_{i-1}\} = \{X^{i-2},Y^{i-2}\}$.  From \eqref{eq:dyn_x} and \eqref{eq:dyn_y} it follows that $X_{i-1}$ and $Y_{i-1}$ can be written as:
\begin{align*}
X_{i-1} &= \alpha^{xx}_1 f_x(X_{i-2}) + \alpha^{zx}_2 f_z(Z_{i-3}) + \alpha^{sx}_1 f_s(S_{i-2}) + W^x_{i-1} \\
&= \alpha^{xx}_1 f_x(X_{i-2}) + \alpha^{zx}_2 f_z(Z_{i-3}) + \alpha^{sx}_1  \beta W_{i-2}^s + W^x_{i-1} \\
&\quad+  o(\alpha). \\
Y_{i-1} &= \alpha^{yy}_1 f_y(Y_{i-2}) + \alpha^{xy}_2 f_x(X_{i-3}) + \alpha^{sx}_1  \beta' W_{i-2}^s \\
&\quad + \alpha^{zy}_2 f_z(Z_{i-3}) +  W^y_{i-1} + o(\alpha). 
\end{align*}
It may be deduced that $\phi = (\alpha^{zx}_2 f_z(Z_{i-3}) +  \alpha^{sx}_1 \beta W_{i-1}^s + W^x_{i-1}, \alpha^{zy}_2 f_z(Z_{i-3}) + \alpha^{sx}_1  \beta' W_{i-2}^s +   W^y_{i-1} )$ is (asymptotically in $\alpha\to 0$) a sufficient statistics for $Z_{i-2}$ given $F'$. Thus, the following conditional Markov chain holds asymptotically:
\begin{align}
    Z_{i-2}|_{F'} - \phi|_{F'} - (X_{i-1},Y_{i-1})|_{F'}.
\end{align}
Consider now $(\alpha^{zx}_2 f_z(Z_{i-4}) +  \alpha^{sx}_1 \beta W_{i-2}^s + W^x_{i-2}, \alpha^{zy}_2 f_z(Z_{i-4}) + \alpha^{sx}_1  \beta' W_{i-3}^s +   W^y_{i-2} )$, which is asymptotically a conditional sufficient statistics for $Z_{i-2}$ given $X_{i-2}$ and $X_{i-2}$. Iteratively applying the above arguments proves that
\begin{align*}
    \bar{\phi} &\triangleq \{\alpha^{zx}_2 f_z(Z_{i-j-2}) +  \alpha^{sx}_1 \beta W_{i-j}^s + W^x_{i-j}, \alpha^{zy}_2 f_z(Z_{i-j-2}) \\
    &\quad + \alpha^{sx}_1  \beta' W_{i-j}^s +   W^y_{i-j} \}_{j=1}^{i-2}
\end{align*}
is a sufficient statistics for $\{X^{i-1}, Y^{i-1}\}$ with respect to $Z_{i-2}$. Thus, for any $c>0$:
\begin{align*}
\lim_{\alpha\to 0} 
I(Z_{i-2} &;  \alpha^{zx}_2 f_z(Z_{i-2}) +  V   |F) \\
&
= \lim_{\alpha\to 0} 
I(Z_{i-2} ;  \alpha^{zx}_2 f_z(Z_{i-2})  +  V   |\bar{\phi}).
\end{align*}

Let us now consider the case, where we condition upon $F$ and $S^{i-2}$. 
Then $\alpha^{zx}_2 f_z(Z_{i-3})  + W^x_{i-1}$ is a conditional sufficient statistics for $X_{i-1}$ with respect to $Z_{i-2}$. It follows that 
\begin{align*}
     \bar{\phi}^s \triangleq \{\alpha^{zx}_2 f_z(Z_{i-j-2}) +    W^x_{i-j}, \alpha^{zy}_2 f_z(Z_{i-j-2})  +   W^y_{i-j} \}_{j=1}^{i-2}
\end{align*}
is a sufficient statistics for $(F, S^{i-2})$ with respect to $Z_{i-2}$. Thus,
\begin{align*}
\lim_{\alpha\to 0} &
I(Z_{i-2} ;  \alpha^{zx}_2  f_z(Z_{i-2})  +  V   |F,S^{i-2}) \\
&= \lim_{\alpha\to 0} 
I(Z_{i-2} ;  \alpha^{zx}_2 f_z(Z_{i-2})  +  V   |\bar{\phi}^s).
\end{align*}

Clearly, for any $c>0, \mathbb{E}[Z_{i-2}| \bar{\phi}^s] < \mathbb{E}[Z_{i-2}| \bar{\phi}]$, i.e., the MMSE due to predicting $Z_{i-2}$ from $\bar{\phi}^s$ is strictly smaller than when predicting from $\bar{\phi}$, since $\bar{\phi}^s$ is a less noisy version of $Z_{i-3}$ than $\bar{\phi}^s$. 
This proves the lemma. 
\end{IEEEproof}

\IEEEtriggeratref{13}

\bibliographystyle{IEEEtran}
\bibliography{sample}

\end{document}